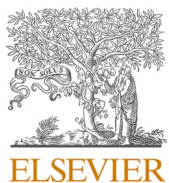

Contents lists available at ScienceDirect

# Polymer

journal homepage: www.elsevier.com/locate/polymer

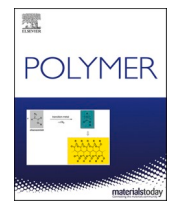

# On the use of neural networks for the structural characterization of polymeric porous materials

Jorge Torre [a,b,c,*], Suset Barroso-Solares [a,b,c], M.A. Rodríguez-Pérez [a,b], Javier Pinto [a,b,c]

[a] *Cellular Materials Laboratory (CellMat), Condensed Matter Physics, Crystallography, and Mineralogy Department, Faculty of Science, University of Valladolid, Spain*
[b] *BioEcoUVA Research Institute on Bioeconomy, University of Valladolid, Spain*
[c] *Study, Preservation, and Recovery of Archaeological, Historical and Environmental Heritage (AHMAT) Research Group, Condensed Matter Physics, Crystallography, and Mineralogy Department, Faculty of Science, University of Valladolid, Spain*

ABSTRACT

The structural characterization is an essential task in the study of porous materials. To achieve reliable results, it requires to evaluate images with hundreds of pores. Current methods require large time amounts and are subjected to human errors and subjectivity. A completely automatic tool would not only speed up the process but also enhance its reliability and reproducibility. Therefore, the main objective of this article is the study of a deep-learning-based technique for the structural characterization of porous materials, through the use of a convolutional neural network. Several fine-tuned Mask R–CNN models are evaluated using different training configurations in four separate datasets each composed of numerous SEM images of diverse polymeric porous materials: closed-pore extruded polystyrene (XPS), polyurethane (PU), and poly(methyl methacrylate) (PMMA), and open-pore PU. Results prove the tool capable of providing very accurate results, equivalent to those achieved by time-consuming manual methods, in a matter of seconds.

## 1. Introduction

Polymeric porous materials are two-phase composites in which a gas phase has been dispersed in a continuous or discontinuous way into a solid phase. Their porous structure provides them with unique properties such as low density, enhanced acoustic and thermal insulation or energy absorption, among others. These properties are known to depend on the characteristics of both the solid matrix and the porous architecture [3]. For instance, pore size can affect the way heat flows through the material, influencing its thermal insulation performance [4,5]. Also, the anisotropy of the pores can induce different mechanical behaviours depending on the direction the material is oriented with respect to the applied forces [5,6]. Moreover, in the biomedical field, polymeric porous materials are being utilized for applications such as tissue engineering [7,8], where their porous architecture provides a suitable environment for pore growth and proliferation, being necessary that the pore size and connectivity meet specific requirements. These materials can also be used as drug delivery systems, with the pore size and structure influencing the controlled release of therapeutic agents [9–11].

All porous structure parameters can be tailored during the manufacturing process to meet specific requirements, which is one of the reasons they are becoming increasingly demanded in numerous sectors [1,12]. Therefore, structural characterization of polymeric porous materials is nowadays a crucial task for both manufacturers and researchers.

Currently, there exists a wide variety of methods for characterizing porous structures [12]. Due to their ease of use, some of the most popular are based on 2D image analysis from micrographs obtained from some kind of microscopy, namely optical [13,14] and scanning/transmission electron microscopy (SEM/TEM) [15,16]. Other techniques such as X-ray tomography offer the possibility of a non-destructive 3D analysis of the porous structure, but generally implying higher costs and time [17,18].

In any case, the user is provided with a set of images that need to be analysed using an image analysis software, i.e., ImageJ [19], QuPath [20], etc., following a specific procedure which is not yet unique and well defined. The closest attempt to stablish a standard image analysis method is the intersections method from the American Society for Testing Materials (ASTM) [21]. However, it proves inconvenient for medium and high-density materials since it does not consider the thickness of the pore walls, thus, resulting in inaccurate outcomes. Also, it cannot provide further information about bimodal porous structures as it can only yield an average of all pore sizes [16]. Another commonly used method is the *manual overlay method* [16]. It consists in manually contouring from 100 to 200 pores per micrograph and evaluating their individual diameters in order to obtain very accurate results for parameters such as the mean two-dimensional pore size ($\overline{\varphi}_{2D}$) or the anisotropy ratio ($R$). This method is independent on the material under study and provides highly reliable results, but requires a large amount of time (around 1 h per 100 pores) and it is significantly subjected to human interpretations and systematic errors. In general, in every method there is a trade-off between achieving accurate results and devoting a large amount of time to it. Therefore, fast, automatic and accurate tools for the structural characterization of porous materials are an actual necessity in today's scientific and industrial outlook.

* Corresponding author. Cellular Materials Laboratory (CellMat), Condensed Matter Physics, Crystallography, and Mineralogy Department, Faculty of Science, University of Valladolid, Spain
*E-mail addresses:* jorge.torre@uva.es (J. Torre), javier.pinto@uva.es (J. Pinto).

At present, only one approach to the development of an automatic tool can be found in the literature and it is the one created by Pinto et al. [16]. They developed a user-interactive tool as a plugin for the software ImageJ [19] that binarizes a micrograph from any source and identifies every pore the user has marked as such. Even though this is an implementation which offers precise results in general, it still has some drawbacks. One of the most notable one is the dependence of the tool's accuracy on the quality of the input images. Although the tool may be designed to work with images from any source, the results may vary depending on the source and the material being analysed, which could hinder a clean binarization. This means that the algorithm must be adjusted differently for each type of image or material, adding to the complexity of the tool's usage. Another shortcoming of this method is that the user must manually point out every single pore that will be analysed using a mouse, which can be a tedious and time-consuming task, and is susceptible to introduce user bias. While the tool can automatically perform all calculations in the background, the need for manual intervention makes it a less efficient process. Finally, it does not provide the user with the masks of the pores, which is an essential aspect of various analytical techniques [17,22].

Instance segmentation is the idea of finding and separating every occurrence or instance of the same class of object in an image by labelling each of its pixels [23–26]. It has become an increasingly studied task in the context of deep learning techniques, namely deep convolutional neural networks (DCNNs), for its capability to solve problems in many fields such as medicine, autonomous driving, security, robotics, etc. [24]. In addition to identifying objects, a new challenge arises when one aims not only to recognize them but also to accurately outline them.

For this purpose, scientists all over the Earth have developed many different DCNN architectures. For example, Mask R–CNN [27] is a relatively simple and flexible DCNN that extends the Faster R–CNN [28] object detection framework to predict object masks in addition to object bounding boxes and class labels. Mask R–CNN achieves state-of-the-art performance on various benchmark datasets while maintaining a time to precision ratio over the average [29]. YOLACT [30] is another state-of-the-art DCNN proposed as a one-stage model which has become very popular for real time instance segmentation due to its high efficiency. However, for some applications, it may sacrifice precision compared to other models [29]. Dual-Swin-L [31], on the other hand, is a more complex DCNN whose architecture is composed of multiple backbones efficiently combined to improve its performance, leading to strong generalization capabilities and, thus, high precision results, albeit at the cost of being slower than most of the other benchmark DCNNs [29]. Also, recently, Meta brought out a shocking example of instance segmentation model: Segment Anything Model (SAM) [32]. This one offers the ability to segment any type of object in an image without the DCNN having trained with that specific object before, a so called zero-shot inference. This model, however, is yet to be further tested and developed to offer the users fine-tuning capabilities.

Among the many examples of DCNN for instance segmentation, this work is only focused on the study of the performance of a Mask R–CNN model, as it achieves time to precision ratios that are over the one achieved by average benchmark models, as well as having plenty of documentation available online, making it simple to train and use. The objective of this article is, therefore, to present and study the performance of a new Mask R–CNN–based tool for the structural characterization of porous materials. The obtained results evidenced the soundness of this approach, achieving an extraordinary performance which could promote this procedure to a gold standard on the field.

## 2. Data and methods

### 2.1. Image acquisition

The dataset used in this study consists of four sets of scanning electron microscopy (SEM) images captured using a FlexSEM 1000 model from Hitachi (Japan). The first three sets depict different polymeric materials, namely XPS (extruded poly-styrene), PU (polyurethane), and PMMA (poly (methyl methacrylate)), all having closed-porous structure. The fourth set of images contains reticulated PU foams images that were kindly offered by Recticel Flexible Foams Inc and have open-porous structure. These porous materials were produced through a diverse range of manufacturing techniques: extrusion foaming for XPS [33], chemical foaming for close and open-pore PU [34], and gas dissolution foaming with $CO_2$ for PMMA [35]. The SEM images were obtained at different magnifications ensuring to capture the wide variety of details of the porous structures, leading to a more feature-based segmentation [36]. Each individual set of images encompasses its own type of microstructures and morphological characteristics, which will help study the performance of the deep learning model at different levels of recognition difficulty.

### 2.2. Dataset annotation and construction

The SEM images utilized in this study were annotated using Fiji (ImageJ) [19], employing the *manual overlay method* [16]. Each pore within the images was manually contoured to ensure accurate delineation (see Fig. 1). This manual annotation approach is widely recognized as the highest reliability reference method for characterizing porous structures in porous materials, with the shortcoming of being very slow (1 h per 100 pores). Subsequently, for each set of images, a standard train-validation split was employed [23,37] (Table 1). Specifically, 80% of the annotated images were allocated for training the Mask R–CNN model, and 20% of the images were reserved for validation. Lastly, six separate images were set aside as the test dataset, to assess the accuracy of the trained model.

### 2.3. Mask R–CNN model structure

In order to perform the challenging task of instance segmentation, Mask R–CNN (Regional Convolutional Neural Network) was used [27]. The overall framework of Mask R–CNN is depicted in Fig. 2, illustrating the flow of information and the main components involved. As a summary, the architecture is composed of an input, a feature extraction section, a detection section, and an output.

As an input, the neural network receives the image in the form of a multi-dimensional array, where each pixel is represented by its height and width coordinates, and the three values corresponding to the RGB color channels. After that, Mask R–CNN operates as a two-stage algorithm. In the first stage, it generates both features and proposals, the latter ones referred to as Regions of Interest (ROIs), around potential object regions in the input image. For this purpose, the stage is composed of two key architectural components: a backbone convolutional neural network that serves as a feature extractor, and a Region Proposal Network (RPN) responsible for generating region proposals.

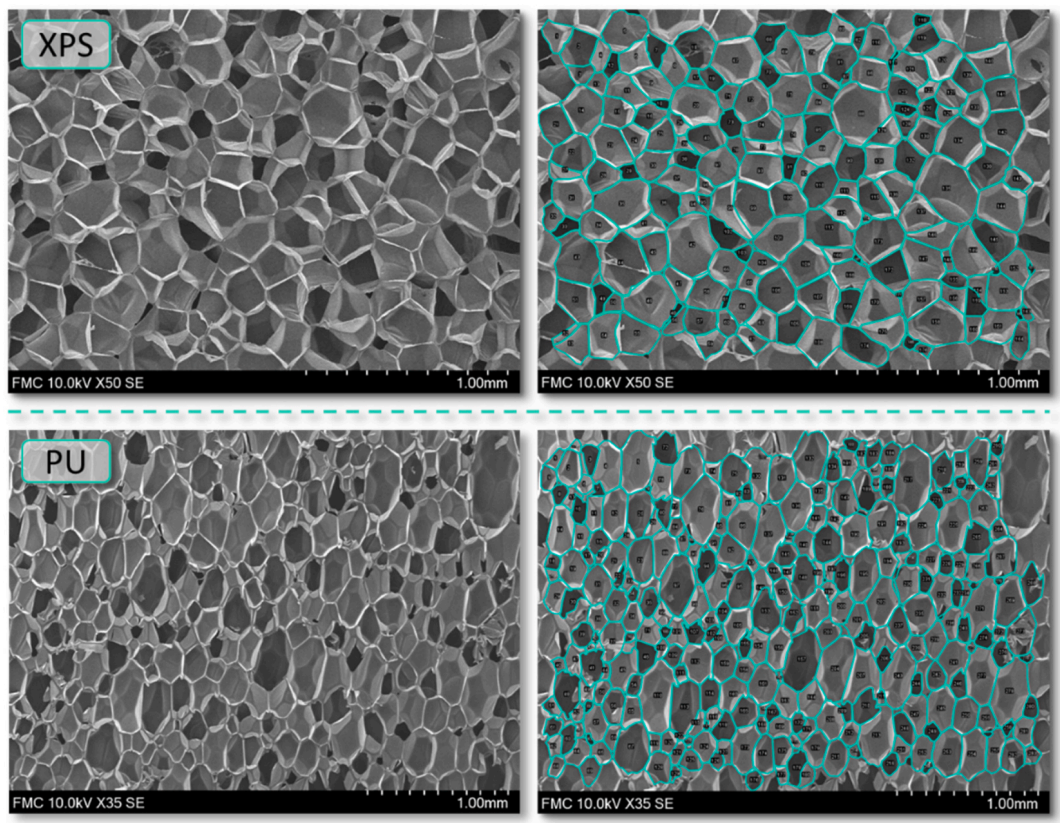

**Fig. 1.** Example of annotation or labeling using the *manual overlay method* in Fiji.

**Table 1**
Characteristics of the four sets of images used to train and validate Mask R–CNN.

| Set | Material | Type of pore | # Images | | # Total pores |
|-----|----------|--------------|----------|------------|---------------|
| | | | Training | Validation | |
| 1 | XPS | Closed | 27 | 7 | 3437 |
| 2 | PU | Closed | 49 | 12 | 4138 |
| 3 | PMMA | Closed | 142 | 36 | 12,511 |
| 4 | PU | Open | 21 | 5 | 1889 |

The Region of Interest (ROI) proposals obtained in the first stage have coordinates represented by real numbers with decimals, given that they are downscaled dividing by the same factor as the feature maps are with respect to the original image (usually a factor of 32). To align both and ensure precise pixel-level alignment without losing any data, the authors of Mask R–CNN introduced the ROIAlign operation. ROIAlign tackles the challenge of accurately sampling and aligning features from each ROI into a *7x7* fixed-size grid feature map [27].

The second stage of Mask R–CNN focuses on refining the bounding box, predicting the object class, and generating a pixel-level mask for each ROI. It consists of two network heads comprising convolutional and fully connected layers. One head is responsible for object classification and bounding box regression, refining the initial proposals. The other head is dedicated to generating precise pixel-level masks for the identified objects [27]. The final output of Mask R–CNN includes the class of the detected objects (in this case only one class named "pore"), the bounding box coordinates for each detected object, and a binary mask that indicates their shape at pixel level.

## 2.4. Mask R–CNN training

### 2.4.1. Machine characteristics

The experiments were conducted on an ASUS TUF GAMING B660M-PLUS WIFI D4 machine, equipped with a 12th Gen Intel(r) Core(TM) i7-12700 processor running at 2100 MHz, featuring 20 logic processors with a total RAM of 128 GB. The main computing element employed was a 16 GB GPU NVIDIA RTX A4000.

### 2.4.2. Transfer learning and fine-tuning

Transfer learning is a powerful technique in deep learning that takes advantage of knowledge gained from pre-trained models on large-scale datasets to solve specific tasks in different domains or with limited training data [23,25,38]. A common approach when performing transfer learning is fine-tuning, where a pre-trained model's weights are adjusted on the target task's data while retaining the knowledge learned from the source domain. During fine-tuning, only the top layers or certain components of the network are updated to suit the target task, while the lower layers, responsible for learning low-level features, maintain their learned representations. This process allows the model to specialize in the target domain, adapt to task-specific characteristics, and achieve improved performance, making transfer learning an indispensable tool.

In this study, fine-tuning was employed to adapt the pre-trained backbone network to the task of instance segmentation in polymeric porous materials. It was done using the pre-trained weights from the widely adopted COCO (Common Objects in Context) dataset [39]. The COCO dataset consists of a vast collection of diverse images spanning 80 object classes. This facilitated the extraction of features relevant to pore identification, while avoiding the need for extensive training from scratch, which would be computationally expensive and data-intensive.

### 2.4.3. Training procedure

In order to gain insights into the model's strengths and limitations in handling the intricate microstructures of porous polymeric materials, the performance of a set of different training configurations was evaluated. For each of the four sets of materials (see Table 1), four distinct trainings were conducted, varying the batch sizes as 1, 2, 8, and 16. In deep learning, *batch size* refers to the number of training examples used in a single iteration of the optimization algorithm during model training. A larger batch size implies that more examples are processed simultaneously, leading to potentially faster convergence, but also increased computational demands. However, the idea that larger batch sizes could lead to worse generalization is still being debated in the scientific community. Currently, the scientific outlook is tending towards the idea that larger batch sizes do not necessarily lead to performance degradation. In fact, increasing the batch size, when combined with an appropriately implemented strategy, can speed up convergence without compromising accuracy [40–44].

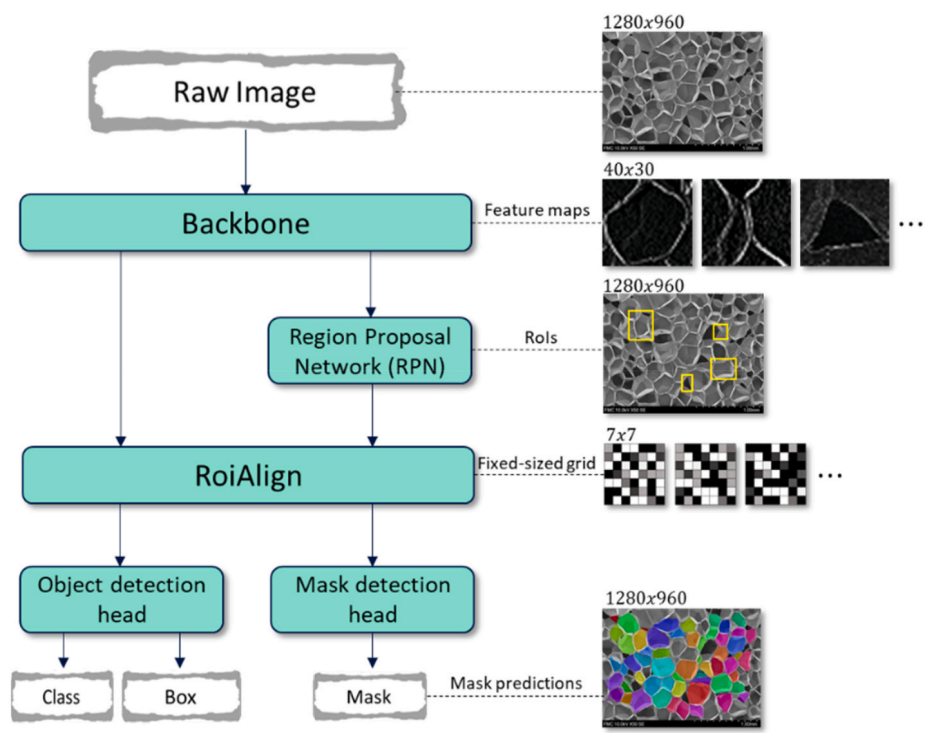

**Fig. 2.** Mask R–CNN instance segmentation framework.

Moreover, the learning rate was set to 0.001, and no learning rate decay was applied throughout the training process. The approach of trying different batch sizes with a fixed learning rate helped investigate the optimal value of both hyperparameters in each intricate dataset. Knowing their optimal values allows for future time optimizations following the rule showcased by Goyal et al. [41]. In their paper, they also show that for batch sizes lower than the training dataset size, there exists an optimal batch size, which will be addressed in this work.

In addition to training each set of materials separately, a combined training where all materials with closed porous structure were merged into a single training dataset was conducted. This aimed to assess the model's performance when exposed to a much wider range of features, thus enabling to investigate potential performance enhancements achieved through collective training.

In each training run, an epoch consisted of a number of steps determined by Equation (1). Image augmentation was deliberately excluded to assess the model's performance under the original data distribution. This decision was made to ensure that the network's segmentation capabilities were primarily reliant on its intrinsic feature learning rather than augmented data representations.

$$\text{steps per epoch} = \frac{\text{\# training examples}}{\text{batch size}} \quad (1)$$

### 2.4.4. Training evaluation metrics

Mask R–CNN uses a "loss function", also called the "objective function", to estimate the quality of predictions made by the network on the training data, for which the actual labels are known. Loss functions are optimized during the learning process of any CNN. They quantify the difference between the estimated output of the model (*prediction*) and the correct output (*ground-truth*) [25,26]. The calculation of the loss function used in a CNN model depends on the end problem. As explained by Fang H. et al. [45], at every epoch end, a total loss (*loss_total*) is calculated for the validation dataset. The cumulative validation loss is the sum of five distinct components, denoted as shown in Equation (2).

$$val\_loss\_total = val\_loss\_rpn\_cls + val\_loss\_rpn\_loc + val\_loss\_cls$$
$$+ val\_loss\_box\_reg + val\_loss\_mask \quad (2)$$

The initial two components, *val_loss_rpn_cls* and *val_loss_rpn_loc*, correspond respectively to the classification loss and localization loss within the regional proposal network (RPN). These metrics assess the performance of the first-stage RPN. Meanwhile, *val_loss_cls* and *val_loss_box_reg* are indicative of losses within the ROI Head, evaluating class labeling and bounding box prediction accuracy, respectively. The equations for this previous losses can be found in the Faster R–CNN original publication [28], as they were inherited in Mask R–CNN. Finally, *val_loss_mask* quantifies the precision of the predicted binary masks generated by the Mask Head, which is the reason why this is the metric that will be used in the document. Mathematically, it is computed as the average binary cross-entropy of the predicted masks pixel distribution and the ground-truth (or manual reference) masks [27,46].

A stopping function was used to stop the training if the validation mask loss explained below did not decrease in 0.001 during at least 10 epochs. This ensures the training is stopped right after the convergence point [25]. However, in some cases with large fluctuations in the test dataset, the stopping function was disabled until a better-performing model in the test dataset was achieved.

### 2.4.5. Model evaluation metrics

The evaluation of the Mask R–CNN model's performance was conducted in two approaches: (I) quantifying the ratio of pores successfully recognized with respect to the total and (II) measuring the accuracy of the generated masks, in both cases taking the *manual overlay method* as a reference.

For both approaches, there exist some standard metrics in instance segmentation evaluation [26,47]. For the first one, precision (*P*) is often used. It measures the fraction of true positive (TP) detections among all positive detections (TP + false positives (FP)). In other words, it quantifies the percentage of pores that where identified correctly among the detected pores. For the second approach, recall (*R*) is a parameter that calculates the fraction of true positive (TP) detections among all actual positive instances (total ground-truth pores). That is, how many pores were identified among the total.

A comprehensive metric that quantifies both of these phenomena is the Mean Average Precision (*mAP*) [26,47]. It is computed by creating a precision-recall curve based on the model's confidence scores. The curve is then interpolated and its integral is computed to yield *mAP*. This is a widely used metric often used to prove state-of-the-art performance. This metric will also be used to provide insights into the model's overall accuracy.

These three parameters only take values from 0 to 1 and are calculated considering a threshold in the Intersection over Union (IoU) metric. This is another parameter that evaluates the accuracy of the generated masks by computing the overlap between the predicted mask and the corresponding ground-truth mask. The results of *P*, *R*, and *mAP* will be given for all detections that fall within at least a threshold IoU value of 0.5.

### 2.5. Porous materials characterization parameters

Let N be the total number of instanced masks using the Mask R–CNN model in a single image. A series of essential parameters were computed to characterize the porous polymeric materials [1,12]. These parameters are considered essential to fully characterize the porous structure and enable to theorize important material characteristics such as thermal conductivity, mechanical resistance, etc. [4–11].

#### 2.5.1. Average pore size ($\overline{\varphi}_{2D}$, $\overline{\varphi}_{3D}$)

The size of each individual pore $\varphi_{2D}^i$ was computed as the average length of four equally spaced diameters originating from the center of its rectangular bounding box. Then, the average pore size $\overline{\varphi}_{2D}$ is the mean value of all of them. Finally, the three-dimensional pore size $\overline{\varphi}_{3D}$ is obtained applying the standard correction factor 1.273 [21], as shown in Equation (3).

$$\overline{\varphi}_{2D} = \frac{\sum_{i=0}^{N}\varphi_{2D}^i}{N}, \overline{\varphi}_{3D} = \overline{\varphi}_{2D}\cdot 1.273 \qquad (3)$$

#### 2.5.2. Anisotropy ratio ($\overline{R}$)

The anisotropy ratio $\overline{R}$ was calculated as the average of the ratios between the largest of four equally-separated diameters $\varphi_{max}$ and its perpendicular $\varphi_{max,\perp}$ for each detected pore $R^i$ as shown in Equation (44).

$$\overline{R} = \frac{\sum_{i=0}^{N}R^i}{N}, R^i = \frac{\varphi_{max}^i}{\varphi_{max,\perp}^i} \qquad (4)$$

#### 2.5.3. Maximum anisotropy angle ($\overline{\theta}_{max}$)

The maximum anisotropy angle $\overline{\theta}_{max}$ was computed as the average angle of the largest diameter of all detected pores (see Equation (5)).

$$\overline{\theta}_{max} = \frac{\sum_{i=0}^{N}\theta_{max}^i}{N} \qquad (5)$$

## 3. Results

### 3.1. Raw dataset: SEM images of polymeric porous materials

In this subsection, the raw data comprising the Scanning Electron Microscope (SEM) images of the four distinct sets of polymeric porous materials in Table 1 is shown. Some examples of these images are showcased in.

Fig. 3. It is important to note that the images are intentionally captured at varying magnifications, ensuring the inclusion of a substantial number of examples and features at different scales.

The SEM images provide a visual representation of the unique characteristics inherent to each material, such as the polygonal-shaped pores of XPS, the rounder shapes of PMMA and closed-pore PU, and the intricate open-pore structure of the reticulated PU.

### 3.2. Closed-pore materials: individual training

#### 3.2.1. Training evaluation: convergence and stability analysis

This subsection presents a study on the convergence and stability of the training process for different batch sizes across each set of closed-pore materials. As this study focuses on the capabilities of the model of producing accurate masks, the validation mask loss (*val_mask_loss*) will be examined as a function of the epoch.

The graphs in Fig. 4 depict the validation mask loss as it evolves over epochs for distinct batch sizes. Notably, all training configurations exhibit a rapid convergence, typically in less than 20 epochs.

In the case of PMMA, while the minimum validation loss is also attained within fewer than 20 epochs, the stopping function was disabled. This is because extending the training to further epochs, such as the 96th epoch in the 1-batch configuration was found to lead to improved results on the test dataset. This, together with the observed fluctuations in the training validation mask loss of PMMA can be attributed to a combination of factors inherent to the material's unique characteristics. Firstly, the training images introduce a challenge due to their occasional high magnification, which often results in image blurring because of the limitations of the SEM microscope. Moreover, the porous PMMA's inherent heterogeneous and intricate nature could potentially hinder the neural network's ability to extract consistent and generalizable features across the entire material. The diversity within PMMA's pore shapes and sizes introduces complexity that might lead to intermittent fluctuations in model performance (Fig. 3).

Additionally, PMMA's high relative density contributes to variations in the appearance of pore walls within the images. In some instances, the thickness of these pore walls is notably larger, resulting in lower brightness along the pore boundaries, thus hindering the recognition of the pore boundaries. Comparing PMMA images with those of PU reveals distinct differences (Fig. 5). PU exhibits more uniform pore structures with clearer brightness along pore borders, facilitated by its lower relative density. This contrasts with the challenges inherent to PMMA, whose denser structure in PMMA results in pore walls that often lack the same level of brightness and well-defined borders found in PU or XPS. Therefore, it is expected that these features contribute to the observed fluctuations in the training validation mask loss. This lack of brightness could be due to the metal deposition prior to the SEM analysis. As the pore size is lower in the PMMA samples, the deposited metal layer's thickness is much more comparable to the pore size, and therefore the amount of metal in the interior could be similar to the one on the walls. In PU, however, much more metal particles could have been deposited within the pores, leaving less amount of it in the walls and, therefore, showing more brightness there. This is, however, a complex question that would require further analysis.

It should also be considered that the test and validation datasets may differ in some specific nuances due to variability and complexity of the PMMA porous microstructure. Initially, the model may have captured a subset of the numerous pore patterns, leading to a rapid decline in validation loss. However, as training progresses, the model refines its understanding of the nuanced features and improves generalization, resulting in better performance on the test dataset.

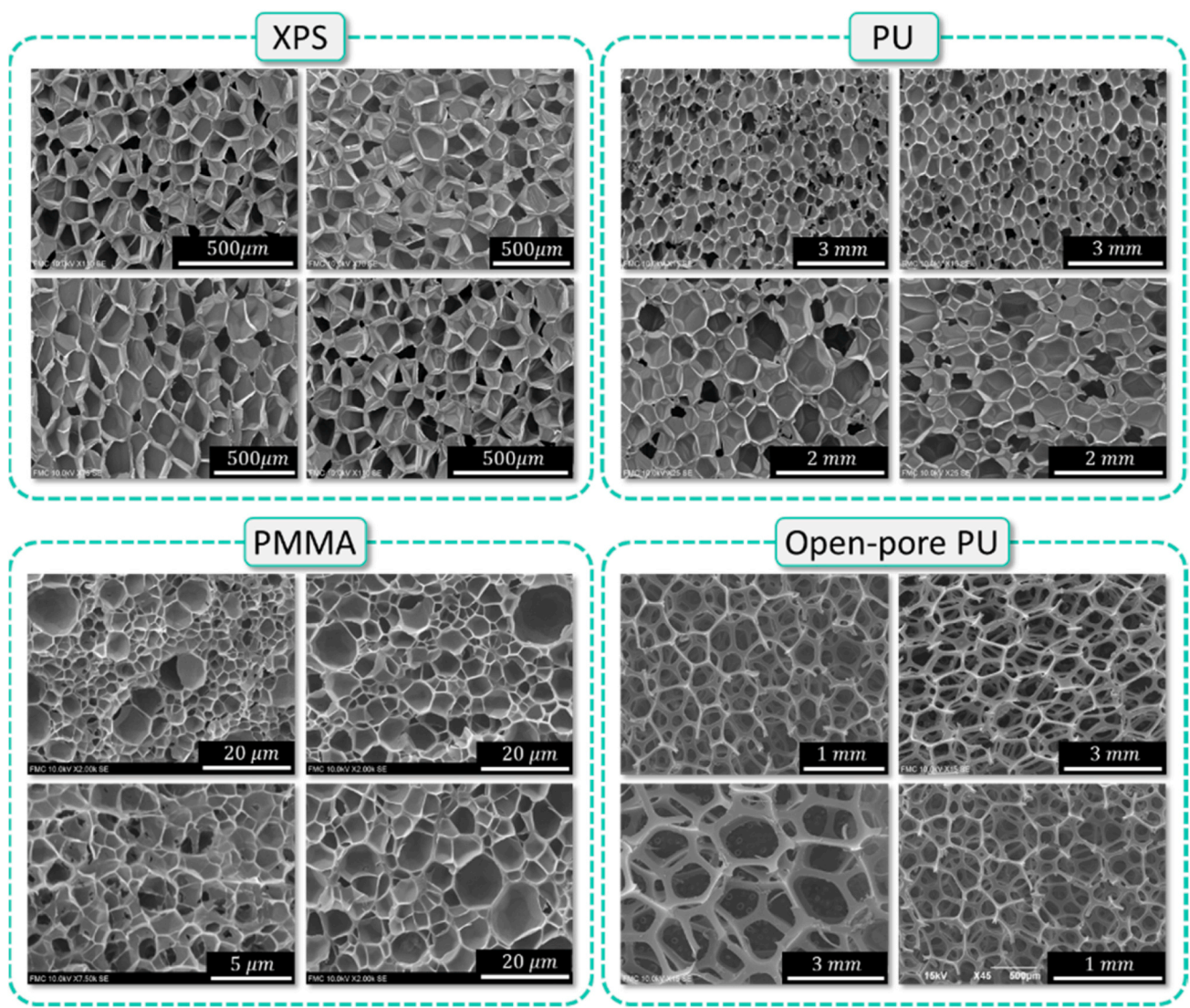

**Fig. 3.** Representative examples of training and validation images for all sets of materials.

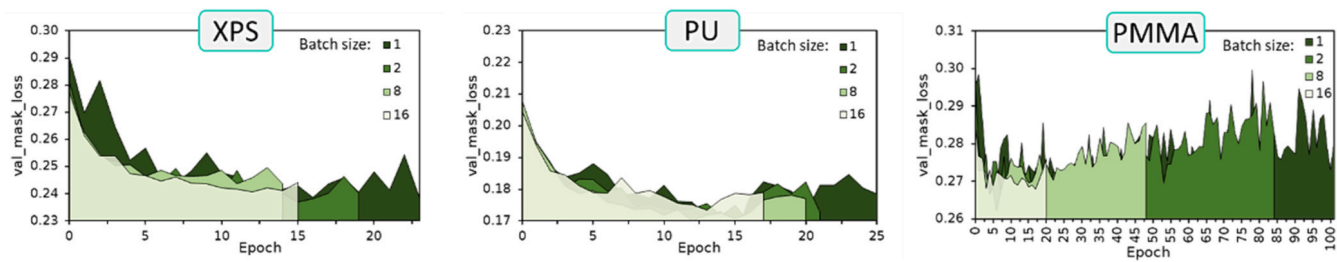

**Fig. 4.** Material-Specific training validation mask loss for XPS, closed-pore PU, and PMMA.

It is important to mention that extending the training beyond the convergence point will, in general, lead to less accurate results, as the model will tend to overfit and, thus, perform worse in external data [25]. This was the case of XPS and PU. All these observations highlight the complicated relationship between the model's accuracy and the complexity of the material, underscoring the need to consider the material-specific complexities when interpreting convergence patterns.

From Fig. 4, it can also be seen that the batch size of 16 stands out as the fastest converging configuration, reaching stable losses within the shortest span of epochs. Also, it can be derived that, as batch size increases, fluctuations in the validation mask loss decrease. This phenomenon suggests that larger batch sizes lead to more stable training dynamics.

### 3.2.2. Models performance evaluation

In Fig. 6, the mask predictions derived from the best-performing batch size configuration for each set of material are shown. In this case, as well as in the entire work, six images from every set of materials

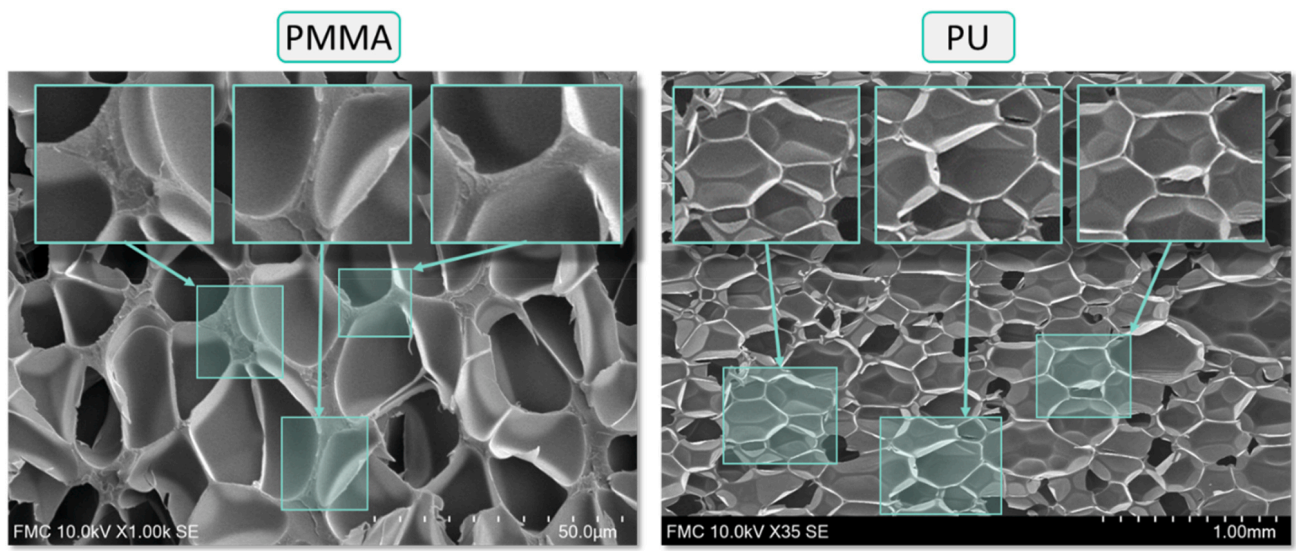

**Fig. 5.** Microstructure Comparison of PMMA and PU Porous Materials. Notably, pore walls have generally better definition in PU due to the increased brightness.

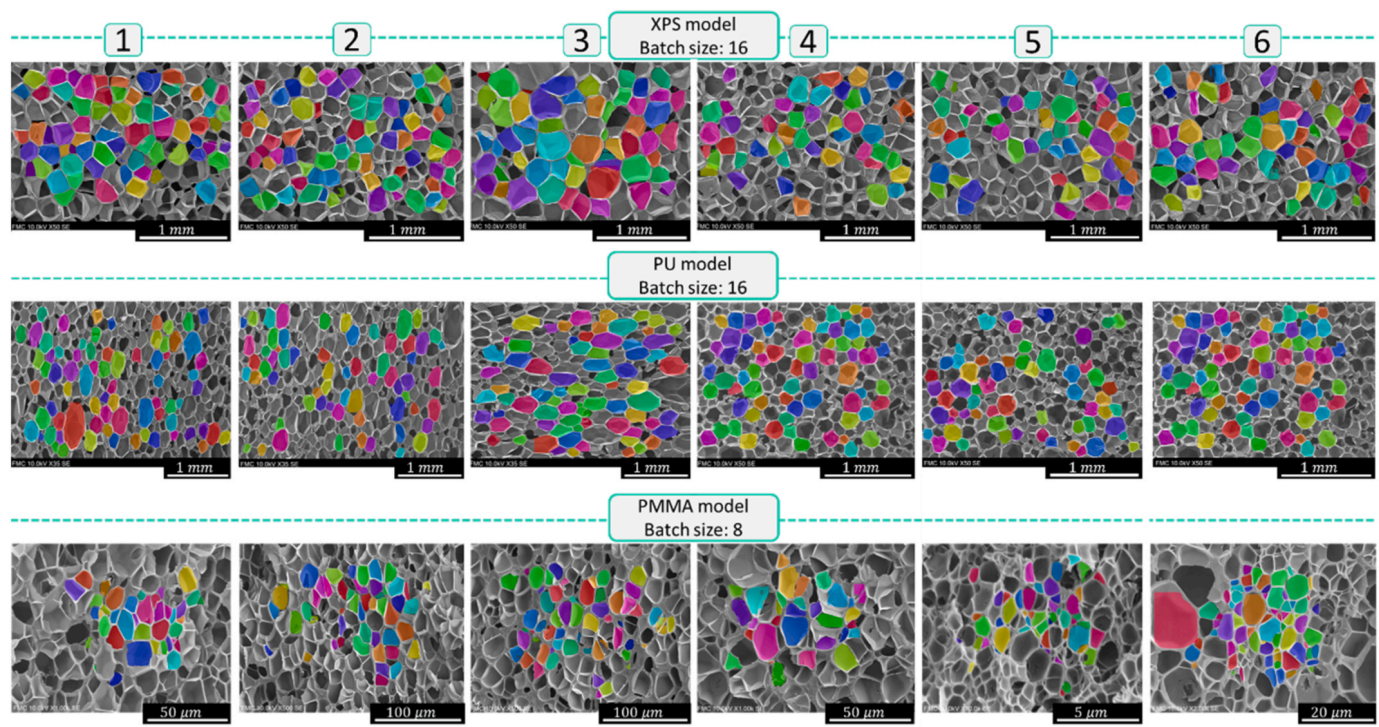

**Fig. 6.** Mask prediction results for the best-performing training configuration of every set of materials. The images correspond to predictions using models trained with the same type of material as in the tested images.

compose the test datasets. Thus, all following results were obtained from these datasets.

The images displayed within the figure provide a representative outlook of the models' capabilities across all materials. It can be seen that the best performing configurations – that is best *mAP*, as it will be quantified later – , were, generally, the ones with larger batch size. This suggests that, for a constant learning rate of 0.001, an optimal value of the batch size has been found.

At a first glance, it can be seen in the images that numerous masks were predicted. The masks were almost always correctly delineated, showcasing the exceptional performance of the models. It can, therefore, be stated that Mask R–CNN is good in understanding what a pore is, even though not every pore in the image is predicted. Also, it becomes apparent again that the most challenging outcomes occur with PMMA. This is something that is clarified with the quantitative results below. Notably, the pores cut by the edges of the images were not delineated, something that is desirable given that, in those cases, the complete shape of the pore is unknown.

The quantitative evaluation of the model's accuracy by means of the aforementioned parameters *mAP*, *P*, and *R* is showcased in Fig. 7. Those values represent the average values of their corresponding metrics over the six test images. It is possible to quantify now that the best-performing models for each material result in a *mAP* of 0.33 in XPS, 0.27 in PU, and 0.25 in PMMA.

Examining the *mAP* results, it can be seen that models tend to exhibit superior performance in materials they were trained on. However, in-

## $mAP$

| mAP - 1Batch | | Trained material | | |
|---|---|---|---|---|
| | | XPS | PU | PMMA |
| Tested material | XPS | 0.23 | 0.28 | 0.21 |
| | PU | 0.01 | 0.15 | 0.08 |
| | PMMA | 0.06 | 0.20 | 0.22 |

| mAP - 2Batch | | Trained material | | |
|---|---|---|---|---|
| | | XPS | PU | PMMA |
| Tested material | XPS | 0.35 | 0.33 | 0.16 |
| | PU | 0.05 | 0.24 | 0.06 |
| | PMMA | 0.10 | 0.19 | 0.13 |

| mAP - 8Batch | | Trained material | | |
|---|---|---|---|---|
| | | XPS | PU | PMMA |
| Tested material | XPS | 0.26 | 0.31 | 0.21 |
| | PU | 0.02 | 0.25 | 0.09 |
| | PMMA | 0.05 | 0.19 | 0.22 |

| mAP - 16Batch | | Trained material | | |
|---|---|---|---|---|
| | | XPS | PU | PMMA |
| Tested material | XPS | 0.37 | 0.36 | 0.10 |
| | PU | 0.04 | 0.28 | 0.04 |
| | PMMA | 0.14 | 0.20 | 0.21 |

## $P$

| P - 1Batch | | Trained material | | |
|---|---|---|---|---|
| | | XPS | PU | PMMA |
| Tested material | XPS | 0.96 | 0.95 | 0.84 |
| | PU | 0.60 | 0.97 | 0.76 |
| | PMMA | 0.74 | 0.94 | 0.91 |

| P - 2Batch | | Trained material | | |
|---|---|---|---|---|
| | | XPS | PU | PMMA |
| Tested material | XPS | 0.98 | 0.95 | 0.90 |
| | PU | 0.85 | 0.97 | 0.82 |
| | PMMA | 0.88 | 0.93 | 0.88 |

| P - 8Batch | | Trained material | | |
|---|---|---|---|---|
| | | XPS | PU | PMMA |
| Tested material | XPS | 0.96 | 0.94 | 0.89 |
| | PU | 0.80 | 0.98 | 0.78 |
| | PMMA | 0.74 | 0.92 | 0.89 |

| P - 16Batch | | Trained material | | |
|---|---|---|---|---|
| | | XPS | PU | PMMA |
| Tested material | XPS | 0.96 | 0.95 | 0.86 |
| | PU | 0.87 | 0.98 | 0.78 |
| | PMMA | 0.90 | 0.93 | 0.88 |

## $R$

| R - 1Batch | | Trained material | | |
|---|---|---|---|---|
| | | XPS | PU | PMMA |
| Tested material | XPS | 0.14 | 0.16 | 0.15 |
| | PU | 0.29 | 0.10 | 0.13 |
| | PMMA | 0.19 | 0.13 | 0.14 |

| R - 2Batch | | Trained material | | |
|---|---|---|---|---|
| | | XPS | PU | PMMA |
| Tested material | XPS | 0.19 | 0.19 | 0.12 |
| | PU | 0.11 | 0.14 | 0.14 |
| | PMMA | 0.13 | 0.12 | 0.11 |

| R - 8Batch | | Trained material | | |
|---|---|---|---|---|
| | | XPS | PU | PMMA |
| Tested material | XPS | 0.15 | 0.18 | 0.14 |
| | PU | 0.16 | 0.14 | 0.11 |
| | PMMA | 0.17 | 0.13 | 0.11 |

| R - 16Batch | | Trained material | | |
|---|---|---|---|---|
| | | XPS | PU | PMMA |
| Tested material | XPS | 0.20 | 0.20 | 0.11 |
| | PU | 0.13 | 0.16 | 0.14 |
| | PMMA | 0.13 | 0.13 | 0.11 |

**Fig. 7.** Results of $mAP$, $P$ and $R$ for every training configuration and for every tested material. The color scale works as a guide for the eyes: it is individual for each metric and takes redder values when closer to the lowest value and greener when closer to the maximum. (For interpretation of the references to color in this figure legend, the reader is referred to the Web version of this article.)

stances also arise where models trained on certain materials excel in recognizing different ones. A notable example is the PU-trained model showcasing better performance in PMMA than the PMMA-trained model. This phenomenon boils down to the previously exposed challenges introduced by factors such as image quality and the intricacies of PMMA's porous structure. The same can be said for XPS, although to a lower extent as shown by the results in Fig. 7. The ability of the models to generalize beyond their training material signifies their potential to capture cross-material commonalities and extend their applicability.

Focusing on the $mAP$ again, the model trained with PU emerges as a standout performer across various materials. This phenomenon can be attributed to the PU's well-defined and homogenous porous structure, characterized by distinct pore borders and consistent brightness, which provides the network with consistent and well-generalized features, adaptable to other materials, resulting in a comparably better performance.

Precision ($P$) emerges as a standout strength across all the models, consistently surpassing 0.9 in instances where the tested material aligns with the training material. This high precision underscores the model's proficiency in accurately delineating pores when it possesses prior knowledge of the material's features. Moreover, evaluating $P$ across different materials reveals that batch size appears to play a role, with larger batch sizes correlating with higher overall precision.

Looking at recall ($R$), significantly lower values are observed, indicating that the model often predicts fewer than 20 % of the total pores present in the images. However, while recall values are generally low, the detected pores exhibit high precision, as reflected by the precision metric. Contrary to expectations, the relationship between recall and training material is not consistent. The models' discernment capabilities vary, with XPS and PU emerging as stronger performers.

### 3.2.3. Porous materials characterization evaluation

The present subsection focuses on the in-depth analysis of the Mask R–CNN models' ability to correctly characterize porous materials.

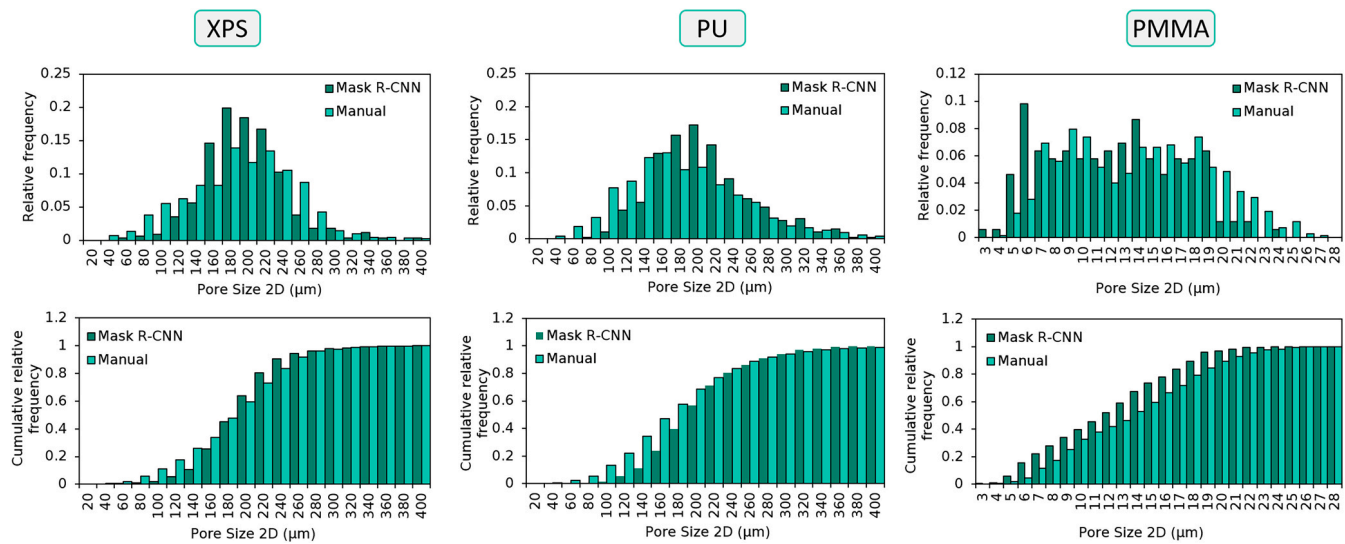

**Fig. 8.** Average pore size distribution comparison and cumulative frequency distribution obtained by the *manual overlay method* and by the best Mask R–CNN training configuration of the models trained with each XPS, PU, and PMMA. The values correspond to tests in materials belonging to the same polymer the models were trained with.

### 3.2.4. Average pore size ($\overline{\varphi}_{2D}$, $\overline{\varphi}_{3D}$)

The histograms in Fig. 8 display the relative frequency distribution of pore sizes in XPS, PU, and PMMA materials, both manually and through the Mask R–CNN method. The accompanying cumulative relative frequency plot provides a comprehensive visualization of the observed pore size distribution patterns across these materials.

In the examination of the pore size histograms of Fig. 8, a notable difference between the Mask R–CNN method and the manual approach is revealed. It can be seen that the Mask R–CNN models tend to omit the smallest pores that the manual method encompasses in the XPS and PU test images. Consequently, this contributes to a higher average pore size calculated with the automatic method (see Fig. 9, PU example). The opposite appears to be happening in PMMA, in which the model had a tendency to take the smallest pores, contributing to a lower average pore size compared to the manual method. This can be visually seen in Fig. 6.

If this is put together with the consistently high precision (P) values achieved by the Mask R–CNN approach exposed in Fig. 7, it can be inferred that this phenomenon of omitting the smallest pores is the main reason explaining the lower average pore size observable in Fig. 9. This is, indeed, further ensured by the examples in Fig. 10, which presents a direct comparison between the pore outlining outcomes achieved through the *manual* method and the Mask R–CNN approach. The illustration exposes some examples in which the smaller pores have not been taken into consideration by Mask R–CNN, as well as some predicted masks that are noticeably smaller than the manual outline. However, the impact extent of the errors made by the model in predicting the pore borders is minimal in general, further proving the overall high precision (P) values evident in Fig. 7.

A key aspect to note, nonetheless, is that even though the magnitude of these errors is small compared to the other error source, their collective impact consistently influences pore size calculations in a downward direction. This is attributed to the model's tendency to predict pores slightly smaller than those outlined manually and not otherwise. To mitigate this issue, a more exhaustive annotation or labeling before training would be necessary, including all pores regardless of their size.

All in all, while differences in pore size results between the two methods may exist, the overall agreement with the manual overlay method, which serves as the reference standard, is evident. The congruence observed underscores the efficacy of the Mask R–CNN approach in providing reliable and consistent outcomes for pore size analysis.

### 3.2.5. Anisotropy ratio ($\overline{R}$) and maximum anisotropy angle ($\overline{\theta}_{max}$)

The calculation of anisotropy was exclusively performed on the test images exhibiting clear anisotropy which, in this study, were solely represented by test images 1–3 corresponding to the closed-pore PU material. Results are showcased in Fig. 11.

In terms of the anisotropy ratio analysis, the comparison between values obtained through both the *manual* and Mask R–CNN methods showcases a remarkable agreement, with a difference of approximately 5–8%, which can be considered marginal. This close alignment between the outcomes affirms the reliability of the Mask R–CNN tool in efficiently and accurately computing anisotropy ratios.

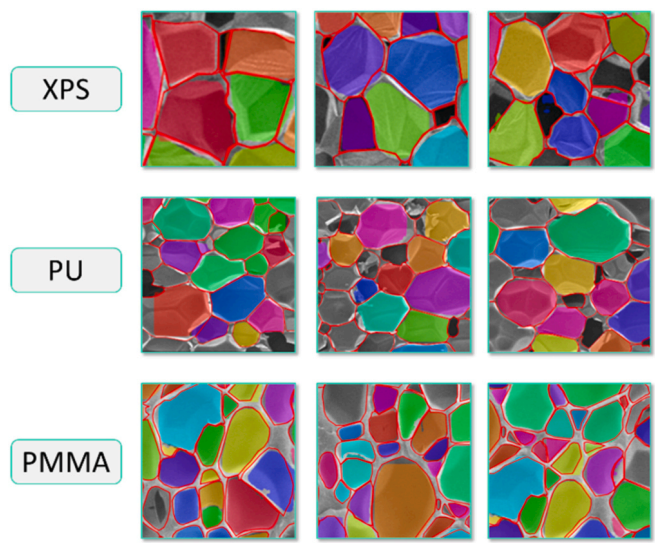

**Fig. 10.** Comparison between pore outlining outcomes: *manual* method vs. individual Mask R–CNN models. The red line represents the *manual overlay method* mask outline; the color-filled masks represent the Mask R–CNN models mask predictions. (For interpretation of the references to color in this figure legend, the reader is referred to the Web version of this article.)

Similarly, when it comes down to determining the maximum anisotropy angle, the findings from the *manual* and Mask R–CNN methods also exhibit accordance. The variance between the two sets of results ranges from 6 to 10°. Therefore, the remarkable consistency observed in both the anisotropy ratio and maximum anisotropy angle analyzes proves the potential of the Mask R–CNN model as a reliable tool for precise anisotropy characterization.

### 3.3. Closed-pore materials: combined training

This section presents an in-depth exploration of a Mask R–CNN model trained on a composite dataset, which is a union of the XPS, closed-pore PU, and PMMA material sets. With this, the aim is to investigate the model's performance in recognizing common as well as distinctive features across different porous structures and assess whether this single model performs better than the sum of the individual ones.

### 3.3.1. Training evaluation: convergence and stability analysis

In this subsection, the dynamics of the training process of the combined model are studied by analyzing the fluctuations in the validation mask loss, analogously to the procedure in the individual models.

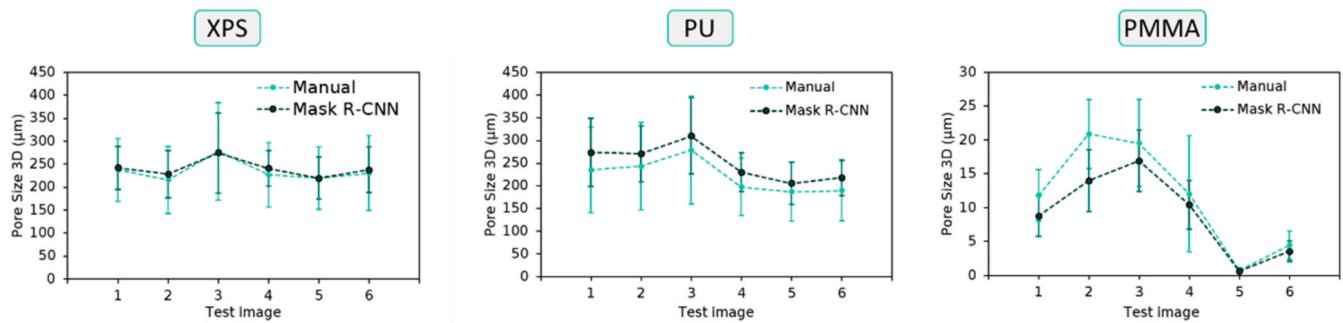

**Fig. 9.** Comparison of pore size 2D results in every test image: *manual* method vs. individual Mask R–CNN models. The vertical bars represent the standard deviation of the calculus of the pore size 3D.

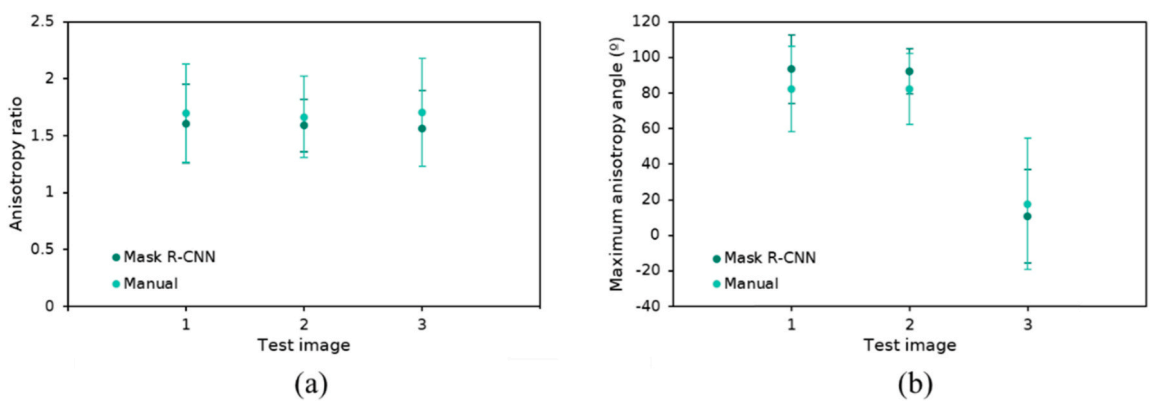

**Fig. 11.** Comparison between outcomes of *manual* method and PU-trained Mask R–CNN model in (a) anisotropy ratio, (b) maximum anisotropy angle for the test images 1–3 of closed-pore PU (see Fig. 6). The vertical bars represent the standard deviation of the calculus of each parameter.

The validation mask loss plots for the combined training of the Mask R–CNN model in Fig. 12 reveal that the convergence in scenario occurs over a comparatively larger number of epochs compared to the individual training configurations. This outcome aligns with expectations, given that the combined training merges three distinct closed-pore datasets and, therefore, the network is exposed to a more extensive set of features and nuances, requiring additional epochs to learn and adapt effectively. Moreover, the observed reduction in fluctuations in validation mask loss with larger batch sizes reaffirms the inherent capability of batch size of stabilizing the training dynamics.

### 3.3.2. Model performance evaluation

The visual results of the best configuration of the combined model are depicted in Fig. 13. Upon initial observation, a greater number of predicted pores is found when compared to the individual models. Notably, the predictions continue to exhibit remarkable precision but remain constrained in terms of recall, albeit to a lower extent. This time, the best performance was obtained with a batch size of 2. In contrast to the trends observed in the individual models, the combined model demonstrates its best performance in smaller batch sizes, in terms of both precision and recall. In addition, trainings with larger batch sizes incurred significantly longer times due to the larger training dataset making the even less viable. These results emphasize the complexity of the task and the necessity for careful consideration when configuring batch sizes for deep learning models.

The analysis of mean Average Precision (*mAP*), precision (*P*), and recall (*R*) metrics in Fig. 14 reveals that the combined model outperforms every single individual model. This outcome aligns with the expectations, as the combination of diverse datasets enables the model to extract common features present in all materials while benefiting from a broader spectrum of them. The concept of training a single neural network with a dataset containing multiple features, as opposed to training individual networks, is supported by several studies [48–50]. Training on multiple datasets helps the model generalize better. It reduces the model's susceptibility to overfitting on particular features or anomalies that might be specific to a single dataset. Besides, training on a larger, more diverse dataset inherently provides more useful data, which can lead to statistically stronger models, leaving aside anomalous data points.

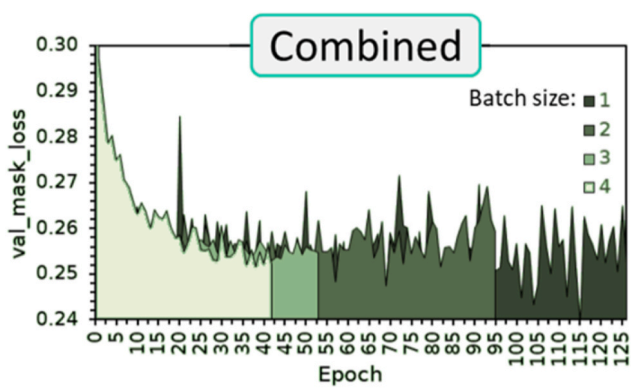

**Fig. 12.** Combined training validation mask loss.

Regarding the best performing model (bath size of 2) from now on, the recall values have reported significant enhancements, reaching 0.26 in XPS, with images having up to 96 accurately identified pores, and 0.19 in PU, with images presenting as many as 122 identified pores. In the case of PMMA, we also observe an improvement in recall (0.17), with one test image revealing up to 80 identified pores. Remarkably, these advancements in recall do not compromise precision, which consistently remains close to the optimal value of 1 in comparison with the individual models.

All of this translates into values of *mAP* of 0.49 in XPS, 0.35 in PU, and 0.28 in PMMA. These results reflect the complexity of the task at hand. Notably, benchmark works like the one from the original authors of Mask R–CNN on the COCO *test-dev* dataset achieve *mAP* values around 0.6, which is indicative of the magnitude of the challenge of using Mask R–CNN in polymeric porous materials. Other studies, such as that by Dangfu Yang et al. [51], showcase impressive *mAP* values close to 1 in their task of segmenting cobble and ballast stones, albeit their images exhibit remarkably better contrast and color conditions. A proof that this can also be attained in porous materials is shown in Fig. 15, where the capabilities of the combined model are put to the test in images of a different PU-foam with better-defined and circular pore shapes. An average recall of 0.88 is obtained, which means that almost the totality of the pores was predicted, save for some black holes, which could have even been easily detected if the model had also been trained with this kind of material. Even so, these results show that the fact that the tool is finding difficulties in more intricate materials is boiled down to Mask R–CNN not being capable of learning every single feature or nuance present such complex images, and not entirely to issues in the training process or wrong hyperparameter settings.

### 3.3.3. Porous materials characterization evaluation

This subsection presents an analysis of the combined Mask R–CNN model performance in the task of correctly characterizing porous materials.

#### 3.3.3.1. Average pore size $(\overline{\varphi}_{2D}, \overline{\varphi}_{3D})$. In Fig. 16, the histograms depicting the relative frequency and combined relative frequency of pore sizes for both the *manual* and Mask R–CNN combined model methods are displayed. A significant improvement is apparent in this instance with respect to the individual models. This time, the histograms

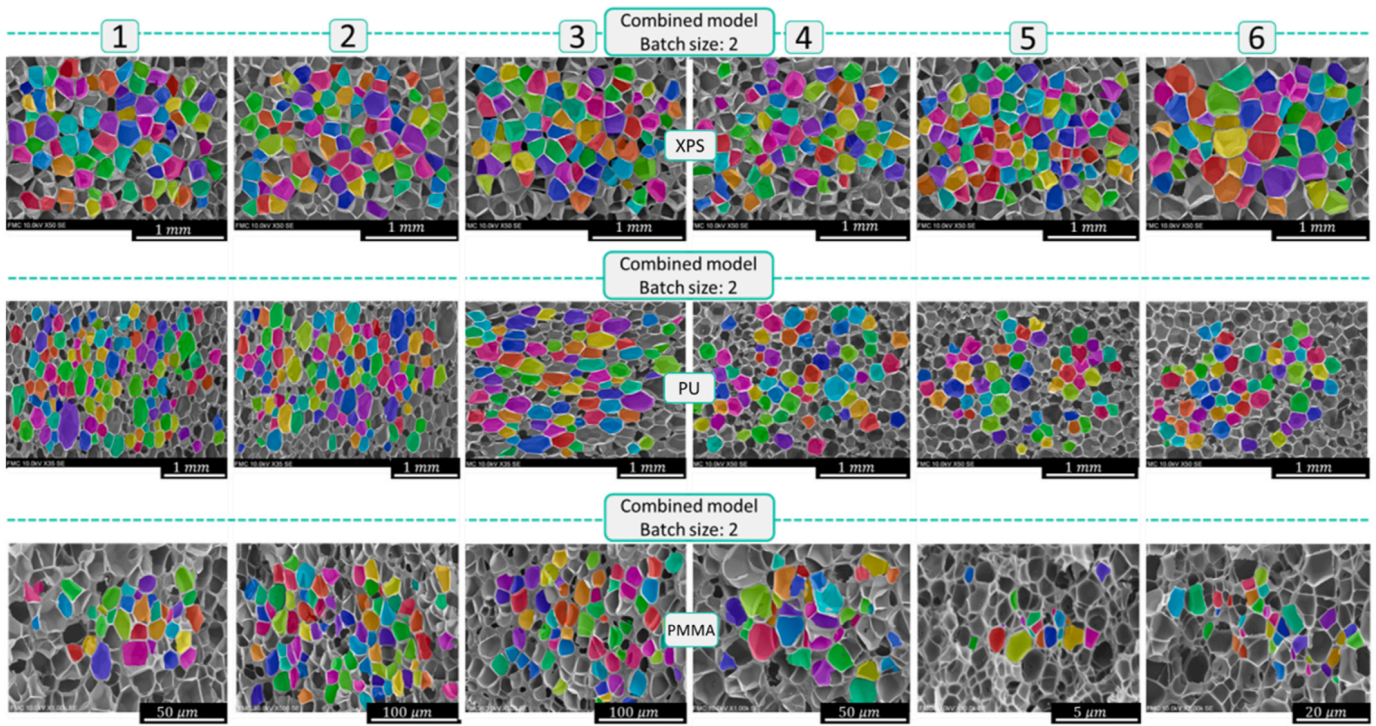

**Fig. 13.** Mask prediction results for the best-performing training configuration of the combined model.

### mAP

| mAP - 1Batch | | Trained material |
|---|---|---|
| | | All |
| Tested material | XPS | 0.38 |
| | PU | 0.24 |
| | PMMA | 0.27 |

| mAP - 2Batch | | Trained material |
|---|---|---|
| | | All |
| Tested material | XPS | 0.49 |
| | PU | 0.35 |
| | PMMA | 0.28 |

| mAP - 8Batch | | Trained material |
|---|---|---|
| | | All |
| Tested material | XPS | 0.21 |
| | PU | 0.07 |
| | PMMA | 0.10 |

| mAP - 16Batch | | Trained material |
|---|---|---|
| | | All |
| Tested material | XPS | 0.19 |
| | PU | 0.10 |
| | PMMA | 0.11 |

### P

| P - 1Batch | | Trained material |
|---|---|---|
| | | All |
| Tested material | XPS | 0.98 |
| | PU | 0.97 |
| | PMMA | 0.93 |

| P - 2Batch | | Trained material |
|---|---|---|
| | | All |
| Tested material | XPS | 0.97 |
| | PU | 0.98 |
| | PMMA | 0.95 |

| P - 8Batch | | Trained material |
|---|---|---|
| | | All |
| Tested material | XPS | 0.97 |
| | PU | 0.94 |
| | PMMA | 0.93 |

| P - 16Batch | | Trained material |
|---|---|---|
| | | All |
| Tested material | XPS | 0.96 |
| | PU | 0.95 |
| | PMMA | 0.93 |

### R

| R - 1Batch | | Trained material |
|---|---|---|
| | | All |
| Tested material | XPS | 0.20 |
| | PU | 0.14 |
| | PMMA | 0.16 |

| R - 2Batch | | Trained material |
|---|---|---|
| | | All |
| Tested material | XPS | 0.26 |
| | PU | 0.19 |
| | PMMA | 0.17 |

| R - 8Batch | | Trained material |
|---|---|---|
| | | All |
| Tested material | XPS | 0.13 |
| | PU | 0.10 |
| | PMMA | 0.11 |

| R - 16Batch | | Trained material |
|---|---|---|
| | | All |
| Tested material | XPS | 0.13 |
| | PU | 0.09 |
| | PMMA | 0.11 |

**Fig. 14.** Results of *mAP*, *P* and *R* of the combined model for every tested material. The color scale uses the same scale and minimum and maximum values as in Fig. 7. (For interpretation of the references to color in this figure legend, the reader is referred to the Web version of this article.)

for the combined model and *manual* method exhibit a significant resemblance to each other. This reinforces the reliability of the combined model, as it successfully considers both small and large pores. Despite this advancement, it remains noteworthy that the combined model still tends overlook the smallest pores in PU when compared with the manual method, further proving the importance of having good-quality training data, that is images with high resolution and well-defined pore borders. However, there has still been an evident improvement when compared with the individual PU model.

Below, Fig. 17 provides a comparison of the pore size 2D results across all test images, comparing the *manual* method against the combined Mask R–CNN method. The outcomes reveal outstanding similarities between the two approaches, underscoring Mask R–CNN's reliability in accurately characterizing pore sizes, despite its inability to detect every pore present in the images. It is important to note that these results are derived from each individual test images, where the manual method has outlined approximately 300 pores while the Mask R–CNN combined model has identified only around $17 - 25\%$ of them, as mirrored by the recall in Fig. 14. However, should the user need to ensure that the results are representative of the whole sample, they could take advantage of the tool's scalability and introduce additional images of that same sample as input, reaching any value needed and obtaining results in a significantly smaller amount of time compared to the *manual* method.

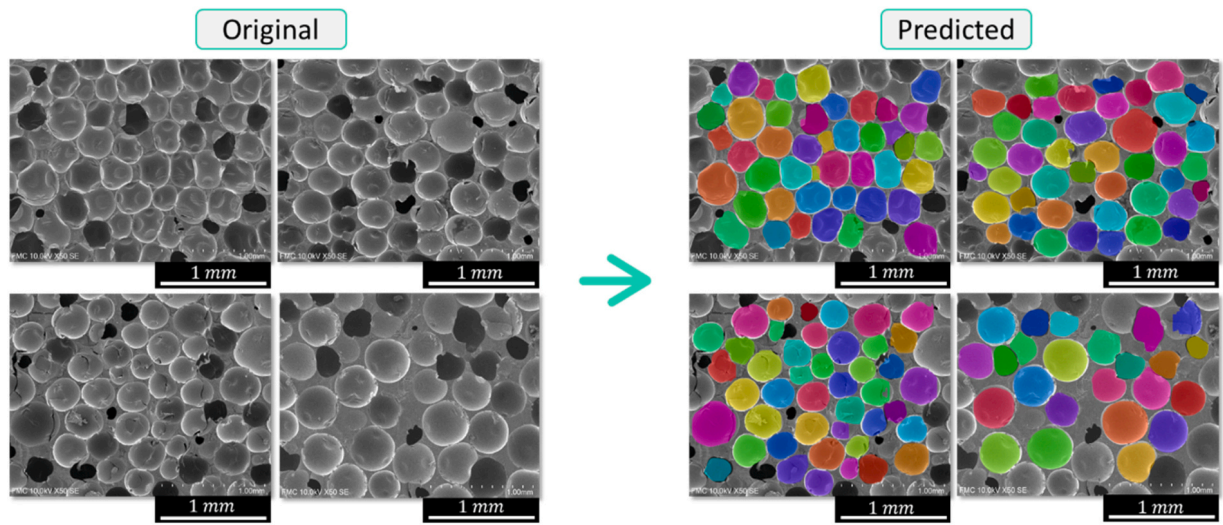

**Fig. 15.** Best performing combined model predictions (batch size of 2) for images of closed-pore PU with well-defined circle-shape pores. An average recall of 0.88 is obtained.

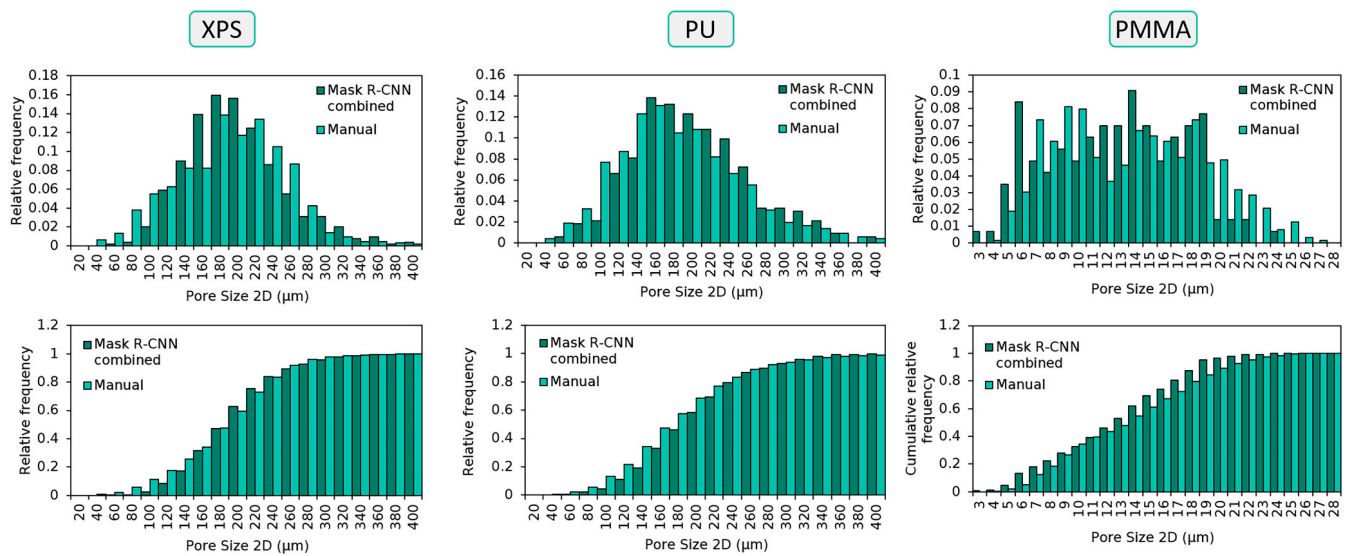

**Fig. 16.** Average pore size 2D distribution comparison and cumulative frequency distribution obtained by the *manual overlay method* and by the best combined Mask R–CNN training configuration with a batch size of 2.

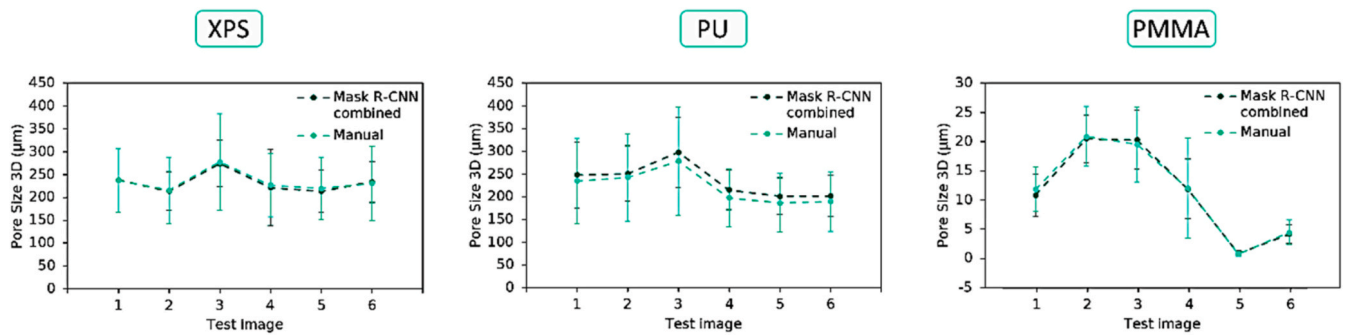

**Fig. 17.** Comparison of pore size 2D results in every test image: *manual method* vs. best performing combined model.

*3.3.3.2. Anisotropy ratio ($\overline{R}$) and maximum anisotropy angle ($\overline{\theta}_{max}$).*
Fig. 18, represents a comparison is between the anisotropy ratio and the maximum anisotropy angle calculated using both the *manual* method and the combined Mask R–CNN model. Once again, the results

demonstrate a closer alignment compared to the individual model. The margin of error is diminished to a range of $0.5 - 5\%$ for anisotropy ratios and $3 - 10$ degrees for the maximum anisotropy angle.

With all these results, the combined model has demonstrated the ability of this Mask-RCNN-based tool to accurately characterize nearly 100 pores per image in about 7 s. This remarkable speed is a significant leap forward compared to the current *manual* method, which typically requires around 1 h to characterize a similar number of pores. This would position this automatic tool as the fastest tool currently available, being around 515 times faster than the most commonly used method at the present. Additionally, even though it was not used during this study, the tool includes a method for user supervision, allowing the quick removal of any wrongly predicted pore, facilitating even more accurate results.

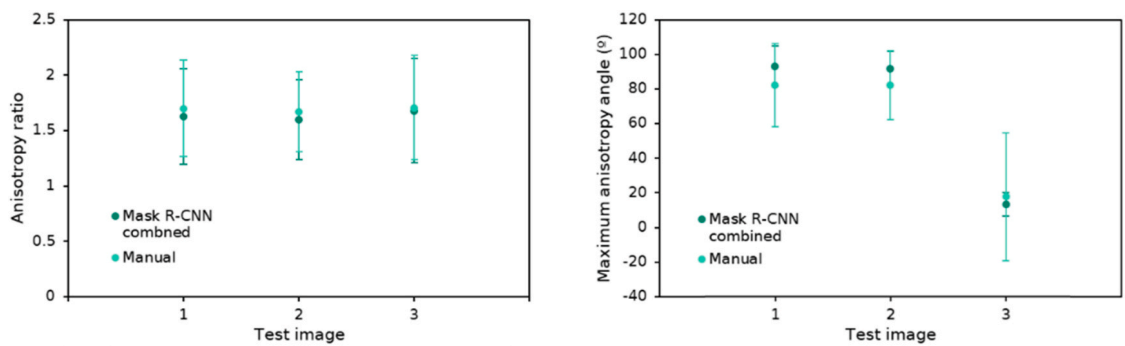

**Fig. 18.** Comparison between outcomes of *manual* method and the combined Mask R–CNN model in (a) anisotropy ratio, (b) maximum anisotropy angle for the test images 1–3 of closed-pore PU (see Fig. 6).

### 3.4. Open-pore material: individual training

This section studies the performance of the convolutional neural network on open-pore PU (see Table 1). Unlike previous sections, the task here differs slightly. Now, Mask R–CNN has been trained to predict the outlines of the pore windows, instead of the outline of the whole pore, as open-pore materials applications often rely on these window sizes for purposes such as gas/liquid filtration or sound/thermal insulation [52–54]. The window size was computed in the same way as the pore size, using the leftmost part of Equation (3).

#### 3.4.1. Training evaluation: convergence and stability analysis

The training evaluation for the open-pore PU material reveals a distinct pattern compared to its closed-pore counterparts. The graph in Fig. 19 illustrating the validation mask loss exhibits notably smoother convergence, indicating a more stable training process. Convergence is again obtained quickly, which is consistent with the closed-pore models. However, it is worth noting that the best results are often attained in subsequent epochs. This phenomenon is attributable to fluctuations in the results on the test dataset, which is separate from the validation dataset. As discussed earlier in the context of closed-pore PMMA, the validation mask loss may not always precisely correlate with the test results due to the differences in the nuances. This behavior emphasizes the importance of examining both training dynamics and actual test performance. It is still worth mentioning that although better results might be achieved, it is not recommended to extend the training further than necessary, as it could fall into overfitting.

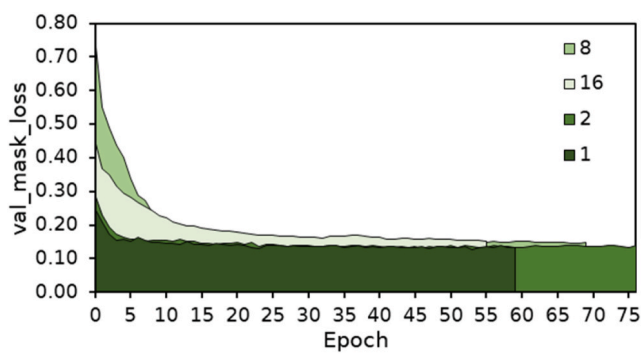

**Fig. 19.** Open-pore PU models training validation mask loss.

#### 3.4.2. Models performance evaluation

Fig. 20 illustrates the mask prediction results obtained from the open-pore PU model's best-performing training configuration. Again, this configuration corresponds to a batch size of 2. This again contrasts with the results observed in the individual models for closed-pore materials. At a first glance, the image exemplifies the model's capability to identify pore window boundaries.

Fig. 21 shows that the open-pore PU model exhibits exceptional recall performance, compared with the results of the previous models, reaching an outstanding value of 0.40 with a batch size of 2. This value contributes to an overall *mAP* of 0.62, which stands out in comparison to the *mAP* results of the closed-pore models. However, it's important to note that this increase in recall is accompanied by a reduction in precision. While the typical precision values for the previous models remain close to 1, in this case, they dip to levels around $0.81 - 0.85$. This could be attributed to the presence of numerous pore struts intersecting above or beneath the pore windows, introducing brighter elements into the images and therefore hindering the correct delineation of the pore boundaries.

#### 3.4.3. Porous materials characterization evaluation

Fig. 22 shows the comparison of the relative frequency histogram and the cumulative relative frequency histogram of the window size distribution obtained through both Mask R–CNN and the manual method for open-pore PU. Notably, both histograms exhibit similar behaviors in the window size in both right and left limits of the distribution, demonstrating consistent consideration of both small and large sizes. However, within the mid-range, some disparities manifest in seemingly random window sizes, but they can likely be attributed to statistical considerations. Importantly, these variations occurring amidst the range are symmetric and therefore compensated, so they are not expected to significantly influence the overall average window size results. This is ensured by Fig. 23, revealing a strong agreement in window size between Mask R–CNN and the *manual* method, reinforcing the robustness and accuracy of the Mask R–CNN technique in open-pore polymeric materials characterization.

The disparities observed in test image number 1 of Fig. 23, where the results obtained through Mask R–CNN yields smaller window size compared to the *manual* method, can be attributed to the fact that, in several instances, Mask R–CNN predicted windows with significantly reduced sizes. This behavior is attributed to the presence of numerous pore struts that overlap and also underlay the pore windows. These struts introduce additional bright pixels within the darker windows, which confuse the neural network during prediction process.

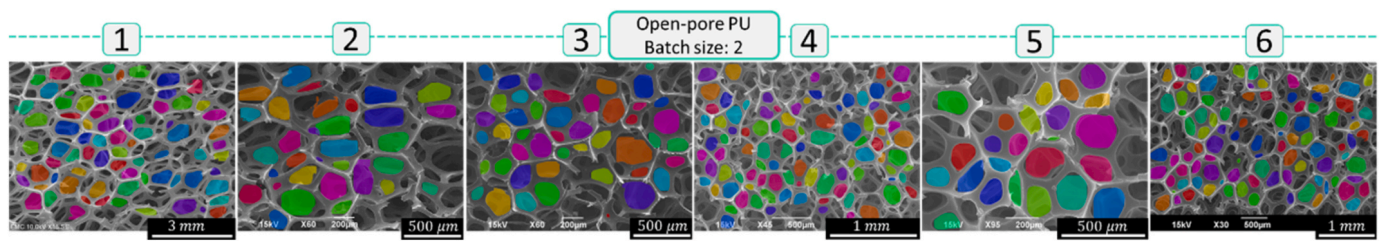

**Fig. 20.** Mask prediction results for the best-performing training configuration of the open-pore PU model.

| mAP | | Trained material |
|---|---|---|
| | | Open-pore PU |
| Batch size | 1 | 0.48 |
| | 2 | 0.62 |
| | 3 | 0.54 |
| | 4 | 0.27 |

| P | | Trained material |
|---|---|---|
| | | Open-pore PU |
| Batch size | 1 | 0.84 |
| | 2 | 0.81 |
| | 3 | 0.85 |
| | 4 | 0.84 |

| R | | Trained material |
|---|---|---|
| | | Open-pore PU |
| Batch size | 1 | 0.29 |
| | 2 | 0.40 |
| | 3 | 0.33 |
| | 4 | 0.19 |

**Fig. 21.** Results of *mAP*, *P* and *R* of the open-pore PU model for every tested material. The color scale uses the same scale as in Fig. 7 and minimum and maximum values of all of them together to enhance comparison with previous results. (For interpretation of the references to color in this figure legend, the reader is referred to the Web version of this article).

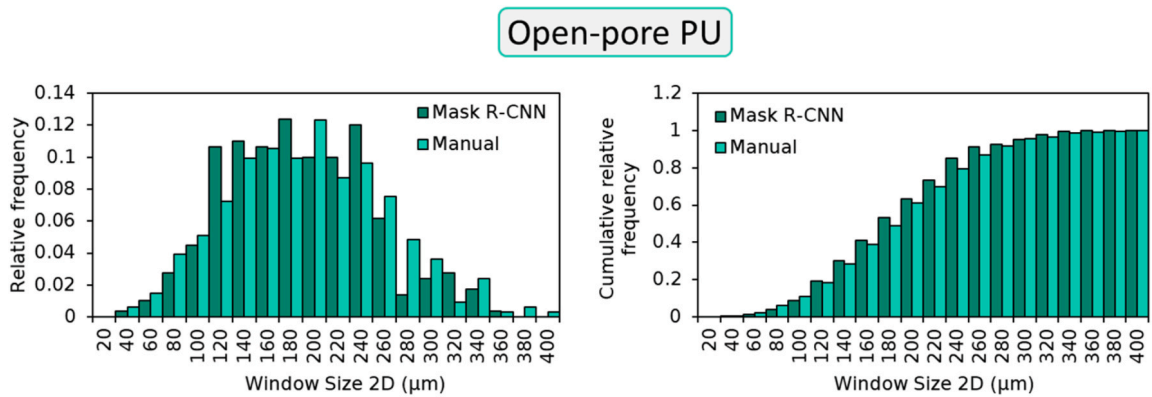

**Fig. 22.** Average pore size distribution comparison and cumulative frequency distribution obtained by the *manual overlay method* and by the Mask R–CNN model trained with open-pore PU and a batch size of 2.

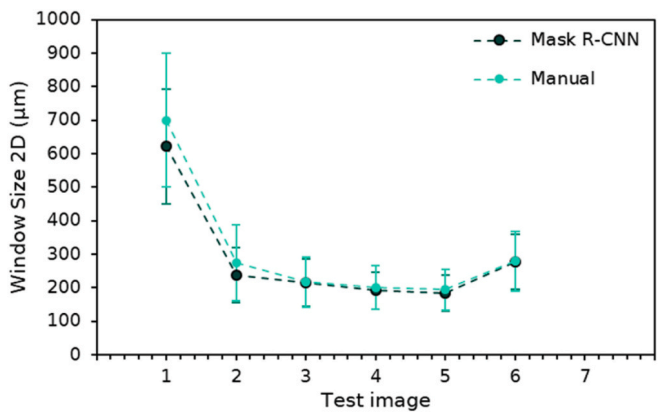

**Fig. 23.** Comparison of window size 2D results in every test image: *manual* method vs. best performing open-PU model.

## 4. Conclusions

In this study, several Mask R–CNN deep learning models have been examined on the task of automatically characterizing the porous structure of porous polymeric materials through SEM microscopic images. The analysis was carried out on three closed-pore materials, namely XPS, PU, and PMMA, and a fourth one with open-pore PU, each presenting distinct structural complexities. For each material, four models with 1, 2, 8, and 16 batch size were trained, as well as a combined model encompassing all closed-pore materials.

The results of the performance evaluations showed that the individual models excelled when tested on the same material they were trained with, although the PU-trained model exhibited outstanding performance when tested on both XPS and PMMA, underscoring the importance of high-quality training images. Moreover, the combined model showcased outstanding performance, overcoming all of the individual models. With this combined model, it was demonstrated that Mask R–CNN can provide highly reliable results in characterizing the pore size, anisotropy ratio, and maximum anisotropy angles, with results equivalent to those of the *manual overlay method*.

Regarding the evaluation of the open-pore PU trained model, it is also seen that the model demonstrates remarkable capabilities in accurately characterizing the window structures. This is because, despite occasional challenges posed by intersecting pore struts, the method achieves accurate masks and predicts large numbers of pore windows, again resembling the *manual* method.

All in all, this study demonstrates that Mask R–CNN offers a promising future for the accurate and efficient characterization of porous polymeric materials. Despite the complexity of the task, the model can effectively characterize diverse porous structures, providing researchers with accurate results in a matter of seconds.

## Tool availability

The tool required to reproduce these findings cannot be shared at this time due to technical or time limitations. However, it will be made public to the user as soon as possible.

## CRediT authorship contribution statement

**Jorge Torre:** Writing – review & editing, Writing – original draft, Visualization, Validation, Software, Project administration, Methodology, Investigation, Formal analysis, Data curation, Conceptualization. **Suset Barroso-Solares:** Writing – review & editing, Supervision, Resources, Investigation. **M.A. Rodríguez-Pérez:** Writing – review & editing, Supervision, Resources, Investigation, Funding acquisition. **Javier Pinto:** Writing – review & editing, Supervision, Resources, Methodology, Investigation, Funding acquisition, Formal analysis.

## Declaration of competing interest

The authors declare the following financial interests/personal relationships which may be considered as potential competing interests:

Míguel Ángel Rodríguez Pérez reports financial support was provided by European Union. Míguel Ángel Rodríguez Pérez reports financial support was provided by Government of Castile and León. Javier Pinto reports financial support provided by the Spanish Government and the European Union. If there are other authors, they declare that they have no known competing financial interests or personal relationships that could have appeared to influence the work reported in this paper.

## Data availability

Data will be made available on request.

## Acknowledgements

Financial assistance from MCIN/AEI/10.13039/501100011033 and the EU NextGenerationEU/PRTR program (PLEC2021-007705), Regional Government of Castilla y León and the EU-FEDER program (CLU-2019-04), and Ministerio de Ciencia e Innovación, EU NextGenerationEU/PRTR program, Regional Government of Castilla y León, and EU-FEDER program "Plan Tractor En Materiales Avanzados Enfocado A Los Sectores Industriales Claves En Castilla Y León: Agroalimentario, Transporte, Energía Y Construcción (MA2TEC)" are gratefully acknowledged.

## References

[1] K. D, K.C. Frisch, Handbook of Polymeric Foams and Foam Technology, Hanser Publishers, 1993, https://doi.org/10.1002/pola.1993.080310535.

[2] C.S. Arroyo, Fabricación De Materiales Celulares Mejorados Basados En Poliolefinas, Relación Procesado-Composición-Estructura-Propiedades, 2013, p. 356, https://doi.org/10.35376/10324/1759.

[3] R.S. Lakes, Cellular solids, J. Biomech. 22 (1989) 397, https://doi.org/10.1016/0021-9290(89)90056-0.

[4] Marcelo J.S. de Lemos, Andreas Öchsner, Graeme E. Murch, Cellular and Porous Materials Thermal Properties Simulation and Prediction, WILEY-VCH, Weinheim, 2020.

[5] Low Density Cellular Plastics, 1994, https://doi.org/10.1007/978-94-011-1256-7.

[6] A.T. Huber, L.J. Gibson, Anisotropy of foams, J. Mater. Sci. 23 (1988) 3031–3040, https://doi.org/10.1007/BF00547486.

[7] S.D. Lucy A Bosworth, Electrospinning for Tissue Regeneration, Woodhead Publishing, 2011, https://doi.org/10.1016/B978-1-84569-741-9.50001-X.

[8] A. Barhoum, Handbook of Nanofibers, Springer, 2019.

[9] C.L. Casper, J.S. Stephens, N.G. Tassi, D.B. Chase, J.F. Rabolt, Controlling surface morphology of electrospun polystyrene fibers: effect of humidity and molecular weight in the electrospinning process, Macromolecules 37 (2004) 573–578, https://doi.org/10.1021/ma0351975.

[10] S. Barroso-Solares, D. Cuadra-Rodriguez, M.L. Rodriguez-Mendez, M.A. Rodríguez-Perez, J. Pinto, A new generation of hollow polymeric microfibers produced by gas dissolution foaming, J. Mater. Chem. B 8 (2020) 8820–8829, https://doi.org/10.1039/d0tb01560a.

[11] A. Vohra, P. Raturi, E. Hussain, Scope of Using Hollow Fibers as a Medium for Drug Delivery, LTD, 2022, https://doi.org/10.1016/B978-0-323-96117-2.00013-3.

[12] S.T. Lee, N.S. Ramesh, Polymer Foams: Mechanisms and Materials, 2004.

[13] Z.M. Ariff, Z. Zakaria, A.A. Bakar, M.A.M. Naser, Effectiveness of A Simple image enhancement method in characterizing polyethylene foam morphology using optical microscopy, Procedia Chem. 19 (2016) 477–484, https://doi.org/10.1016/j.proche.2016.03.041.

[14] J. Kuhn, H.P. Ebert, M.C. Arduini-Schuster, D. Büttner, J. Fricke, Thermal transport in polystyrene and polyurethane foam insulations, Int. J. Heat Mass Tran. 35 (1992) 1795–1801, https://doi.org/10.1016/0017-9310(92)90150-Q.

[15] V. Kumar, N.P. Suh, A process for making microcellular thermoplastic parts, Polym. Eng. Sci. 30 (1990) 1323–1329, https://doi.org/10.1002/pen.760302010.

[16] J. Pinto, E. Solórzano, M.A. Rodríguez-Perez, J.A. De Saja, Characterization of the cellular structure based on user-interactive image analysis procedures, J. Cell. Plast. 49 (2013) 555–575, https://doi.org/10.1177/0021955X13503847.

[17] P. Cimavilla-Román, S. Pérez-Tamarit, S. Barroso-Solares, J. Pinto, M.A. Rodríguez-Pérez, Sub-pixel tomographic methods for characterizing the solid architecture of foams, Microsc. Microanal. 28 (2022) 689–700, https://doi.org/10.1017/S1431927622000447.

[18] S. Pérez-Tamarit, E. Solórzano, A. Hilger, I. Manke, M.A. Rodríguez-Pérez, Multi-scale tomographic analysis of polymeric foams: a detailed study of the cellular structure, Eur. Polym. J. 109 (2018) 169–178, https://doi.org/10.1016/j.eurpolymj.2018.09.047.

[19] M.D. Abràmoff, P.J. Magalhães, S.J. Ram, Image processing with imageJ, Biophot. Int. 11 (2004) 36–41, https://doi.org/10.1201/9781420005615.ax4.

[20] P. Bankhead, M.B. Loughrey, J.A. Fernández, Y. Dombrowski, D.G. McArt, P. D. Dunne, S. McQuaid, R.T. Gray, L.J. Murray, H.G. Coleman, J.A. James, M. Salto-Tellez, P.W. Hamilton, QuPath, Open source software for digital pathology image analysis, Sci. Rep. 7 (2017) 1–7, https://doi.org/10.1038/s41598-017-17204-5.

[21] ASTM D3576-04(2010), Standard Test Method for Cell Size of Rigid Cellular Plastics, 2010, https://www.astm.org/d3576-04r10.html. (Accessed 20 April 2023).

[22] S. Pardo-Alonso, E. Solórzano, L. Brabant, P. Vanderniepen, M. Dierick, L. Van Hoorebeke, M.A. Rodríguez-Pérez, 3D Analysis of the progressive modification of the cellular architecture in polyurethane nanocomposite foams via X-ray microtomography, Eur. Polym. J. 49 (2013) 999–1006, https://doi.org/10.1016/j.eurpolymj.2013.01.005.

[23] I. Zafar, G. Tzanidou, R. Burton, N. Patel, L. Araujo, Hands-On Convolutional Neural Networks with TensorFlow, Packt Publishing, 2018.

[24] A. Mueed Hafiz, G. Mohiuddin Bhat, A survey on instance segmentation, Int. J. Multimed. Inf. Retr. 9 (2020) 171–189.

[25] S. Khan, H. Rahmani, S.A.A. Shah, M. Bennamoun, A Guide to Convolutional Neural Networks for Computer Vision, 2018, https://doi.org/10.2200/s00822ed1v01y201712cov015.

[26] R. StephenShanmugamani, Deep Learning for Computer Vision Expert Techniques to Train Advanced Neural Networks Using TensorFlow and Keras, PacktPublishing, 2018.

[27] K. He, G. Gkioxari, P. Dollár, R. Girshick, R.-C.N.N. Mask, in: Proc. IEEE Int. Conf. Comput. Vis., 2017, pp. 2961–2969.

[28] S. Ren, K. He, R. Girshick, J. Sun, R.-C.N.N. Faster, Towards real-time object detection with region proposal networks, IEEE Trans. Pattern Anal. Mach. Intell. 39 (2017) 1137–1149, https://doi.org/10.1109/TPAMI.2016.2577031.

[29] S. Jung, H. Heo, S. Park, S.U. Jung, K. Lee, Benchmarking deep learning models for instance segmentation, Appl. Sci. 12 (2022) 1–25, https://doi.org/10.3390/app12178856.

[30] D. Bolya, C. Zhou, F. Xiao, Y.J. Lee, YOLACT++ better real-time instance segmentation, IEEE Trans. Pattern Anal. Mach. Intell. 44 (2022) 1108–1121, https://doi.org/10.1109/TPAMI.2020.3014297.

[31] T. Liang, X. Chu, Y. Liu, Y. Wang, Z. Tang, W. Chu, J. Chen, H. Ling, CBNet: a composite backbone network architecture for object detection, IEEE Trans. Image Process. 31 (2022) 6893–6906, https://doi.org/10.1109/TIP.2022.3216771.

[32] A. Kirillov, E. Mintun, N. Ravi, H. Mao, C. Rolland, L. Gustafson, T. Xiao, S. Whitehead, A.C. Berg, W.-Y. Lo, P. Dollár, R. Girshick, Segment Anything, 2023. http://arxiv.org/abs/2304.02643.

[33] C.V. Vo, F. Bunge, J. Duffy, L. Hood, Advances in thermal insulation of extruded polystyrene foams, Cell. Polym. 30 (2011) 137–156, https://doi.org/10.1177/026248931103000303.

[34] F.M. De Souza, M.R. Sulaiman, R.K. Gupta, Materials and chemistry of polyurethanes, ACS Symp. Ser. 1399 (2021) 1–36, https://doi.org/10.1021/bk-2021-1399.ch001.

[35] H. Ventura, L. Sorrentino, E. Laguna-Gutierrez, M.A. Rodriguez-Perez, M. Ardanuy, Gas dissolution foaming as a novel approach for the production of lightweight

biocomposites of PHB/natural fibre fabrics, Polymers 10 (2018), https://doi.org/10.3390/polym10030249.

[36] A. Matskevych, A. Wolny, C. Pape, A. Kreshuk, From shallow to deep: exploiting feature-based classifiers for domain adaptation in semantic segmentation, Front. Comput. Sci. 4 (2022) 1–13, https://doi.org/10.3389/fcomp.2022.805166.

[37] A. Gholamy, V. Kreinovich, O. Kosheleva, Why 70/30 or 80/20 relation between training and testing sets : a pedagogical explanation, Dep. Tech. Reports 1209 (2018) 1–6.

[38] M. Iman, H.R. Arabnia, K. Rasheed, A review of deep transfer learning and recent advancements, Technologies 11 (2023) 1–14, https://doi.org/10.3390/technologies11020040.

[39] Tsung-Yi Lin, Maire M, Belongie SJ, Bourdev LD, Girshick RB, Hays J, Microsoft COCO (Common Objects in Context) dataset, (n.d.). https://doi.org/https://doi.org/10.48550/arXiv.1405.0312.

[40] E. Hoffer, I. Hubara, D. Soudry, Train longer, generalize better: closing the generalization gap in large batch training of neural networks, Adv. Neural Inf. Process. Syst. 2017-Decem (2017) 1732–1742.

[41] P. Goyal, P. Dollár, R. Girshick, P. Noordhuis, L. Wesolowski, A. Kyrola, A. Tulloch, Y. Jia, K. He, Accurate, Large Minibatch SGD: Training ImageNet in 1 Hour, 2017. http://arxiv.org/abs/1706.02677.

[42] C.J. Shallue, J. Lee, J. Antognini, J. Sohl-Dickstein, R. Frostig, G.E. Dahl, Measuring the effects of data parallelism on neural network training, J. Mach. Learn. Res. 20 (2019) 1–49.

[43] S.L. Smith, P.-J. Kindermans, C. Ying, Q.V. Le, Don't Decay the Learning Rate, Increase the Batch Size, 2017. http://arxiv.org/abs/1711.00489.

[44] N.S. Keskar, D. Mudigere, J. Nocedal, M. Smelyanskiy, P.T.P. Tang, On Large-Batch Training for Deep Learning: Generalization Gap and Sharp Minima, 2016. http://arxiv.org/abs/1609.04836.

[45] H. Fang, E. Hovad, Y. Zhang, D. Juul Jensen, Application of Mask R-CNN for lab-based X-ray diffraction contrast tomography, Mater. Char. 201 (2023) 112983, https://doi.org/10.1016/j.matchar.2023.112983.

[46] T. Wang, K. Zhang, W. Zhang, R. Wang, S. Wan, Y. Rao, Z. Jiang, L. Gu, Tea picking point detection and location based on Mask-RCNN, Inf. Process. Agric. 10 (2023) 267–275, https://doi.org/10.1016/j.inpa.2021.12.004.

[47] H. M, S. M.N, A review on evaluation metrics for data classification evaluations, Int. J. Data Min. Knowl. Manag. Process. 5 (2015) 1–11, https://doi.org/10.5121/ijdkp.2015.5201.

[48] M. Ćirić, B. Predić, D. Stojanović, I. Ćirić, Single and multiple separate LSTM neural networks for multiple output feature purchase prediction, Electron 12 (2023), https://doi.org/10.3390/electronics12122616.

[49] J. Xu, Y. Chen, Y. Qin, R. Huang, Q. Zheng, A feature combination-based graph convolutional neural network model for relation extraction, Symmetry 13 (2021) 1–13, https://doi.org/10.3390/sym13081458.

[50] A. Banitalebi-Dehkordi, X. Kang, Y. Zhang, Model composition: can multiple neural networks Be combined into a single network using only unlabeled data?, 32nd, Br. Mach. Vis. Conf. BMVC 2021 (2021).

[51] D. Yang, X. Wang, H. Zhang, Z. yu Yin, D. Su, J. Xu, A Mask R-CNN based particle identification for quantitative shape evaluation of granular materials, Powder Technol. 392 (2021) 296–305, https://doi.org/10.1016/j.powtec.2021.07.005.

[52] C. Zhang, J. Li, Z. Hu, F. Zhu, Y. Huang, Correlation between the acoustic and porous cell morphology of polyurethane foam: effect of interconnected porosity, Mater. Des. 41 (2012) 319–325, https://doi.org/10.1016/j.matdes.2012.04.031.

[53] S. Gunashekar, K.M. Pillai, B.C. Church, N.H. Abu-Zahra, Liquid flow in polyurethane foams for filtration applications: a study on their characterization and permeability estimation, J. Porous Mater. 22 (2015) 749–759, https://doi.org/10.1007/s10934-015-9948-2.

[54] R. Hasanzadeh, T. Azdast, P.C. Lee, C.B. Park, A review of the state-of-the-art on thermal insulation performance of polymeric foams, Therm. Sci. Eng. Prog. 41 (2023), 101808, https://doi.org/10.1016/j.tsep.2023.101808.